# Acquisitive Crimes, Time of Day, and Multiunit Housing in the City of Milwaukee

Scott W. Hegerty
Department of Economics
Northeastern Illinois University
Chicago, IL 60625
S-Hegerty@neiu.edu

## ABSTRACT

According to "Social Disorganization" theory, criminal activity increases if the societal institutions that might be responsible for maintaining order are weakened. Do large apartment buildings, which often have fairly transient populations and low levels of community involvement, have disproportionately high rates of crime? Do these rates differ during the daytime or nighttime, depending when residents are present, or away from their property? This study examines four types of "acquisitive" crime in Milwaukee during 2014. Overall, nighttime crimes are shown to be more dispersed than daytime crimes. A spatial regression estimation finds that the density of multiunit housing is positively related to all types of crime except burglaries, but not for all times of day. Daytime robberies, in particular, increase as the density of multiunit housing increases.

## I. Introduction

While recent U.S. crime rates are recently much lower compared to previous decades, not all central cities have enjoyed this phenomenon. This is particularly true for those whose industrial bases have declined due to globalization. Although a number of theories have been examined to explain these trends, much of the drivers of crime are tied to the strength, or lack thereof, of community bonds and social networks. According to the Social Disorganization theory of criminal behavior, explained by Sampson (1985) and others, crime increases if these networks are weak and if there is a lack of social control to oversee the safety of a neighborhood. This theory, as well as a number of widely tested alternatives, is explained by Hipp (2007) in his study of neighborhood stability and crime in 19 U.S. cities. These alternative theories can be complementary: Smith *et al.* (2000) note the interrelationships and commonalities between Social Disorganization and Routine Activities theory (which focuses on the interactions of perpetrators and potential victims in the absence of capable guardians), for example.

In particular, the type of housing in a neighborhood, and its occupancy, can play an important role in the attraction or deterrence of crime. Higher proportions of renters and population density can both increase crime, but another important factor is the time of day during which residents are present. Traditional distinctions between “daytime,” during which residents are assumed to be at work, and “nighttime,” during which they might be sleeping at home, can vary by income and class. Second- or third-shift service workers might simultaneously present different degrees of attractiveness to property crime and be vulnerable at different times than office workers, for example. It is therefore important to separate differences in income and other socioeconomic variables, as well as times of day during which crimes are most likely to occur, when targeting specific policies in particular neighborhoods. This study attempts to do so, in the context on an analysis of the connection between large rental units and acquisitive crime in one particular American city.

*I.1. A Look at the Literature*

Instability in housing tenure is often recognized as a main contributor to social disorganization, because moving frequently prevents a resident's integration into neighborhood networks. Likewise, renting rather than owning gives residents less of a stake in the stability and long-term success of the surrounding area. This might differ by type of crime. Roncek *et al.* (1981), for example, find that residential blocks near public housing in Cleveland have higher violent crime, but not higher property crime. Likewise, McNulty and Holloway (2000) find proximity to public housing to play an important role in determining violent, but not property, crime in Atlanta.

This phenomenon is not restricted to public housing, however, particularly for property crime. Ceccato *et al.* (2002) show that multifamily housing in Stockholm is related to higher levels to theft of and from cars, as well as of residential burglary; the authors attributed these crimes to an absence of capable guardianship. Lockwood (2007) finds the proportion of renters to be a significant predictor of robberies, but not other crimes. On the other hand, Hipp (2007) finds little connection between burglary or motor vehicle theft and neighborhood stability and poverty, while Raleigh and Galster (2014) find all types of crime in Detroit to be connected with renter occupancy.

An additional factor driving the link between multiunit housing and crime is the fact that these buildings are often occupied differently during different times of day, which might have a varying impact on criminal activity. If an apartment building is relatively vacant during the day, it might be an easier target for thefts, for example. Or, if people are more likely to be home at night, potential perpetrators and victims might be brought into closer contact during this time. But, since "daytime" working hours are by no means universal—particularly among service workers—these patterns might differ by neighborhood. As a result, crime rates might vary by category, by time of day, by income, and by area.

There is relatively little literature regarding these differences, and much of it covers countries outside the United States. Coupe and Blake (2006) offer different underlying motivations, including economic factors, between daytime and nighttime robberies in Britain. Ceccato and Oberwittler (2008) find little difference between the daytime and nighttime models for robbery in Cologne, Germany and Tallinn, Estonia. Montoya *et al.* (2016) find differences in time of day for burglaries in a Dutch city, noting that "target-hardening" is particularly important during the day.

The United States differs from Europe not only in its higher levels of crime, but also in its allocation of housing. Public housing is becoming rarer, even for the very poor; instead, vouchers are provided to support private-market rentals. In addition, since wealthier residents generally live either close to downtown or in suburbs, poorer renters generally live within a certain distance of the city center. These neighborhoods might experience unique variations in crime rates that do not conform to the European literature.

This study examines both the location of large apartment buildings and the time of day for the occurrences of four types of acquisitive crime in the large U.S. city of Milwaukee during 2014. These crimes range from the more personal (Robbery) to less (Theft from Motor Vehicle), with Burglary and Theft of Motor Vehicle likely to be somewhere in between. We combine two strands of descriptive analysis: the categories, times, and spatial patterns of acquisitive crime; and the classification, location, and spatial distributions of apartment buildings and other multiunit housing. We do this by first examining distances between reported crimes and apartment buildings, before estimating a model of per-capita crime by block group, which includes the density of these buildings.

Overall, we find differences among both the time and type of crime. In particular, Thefts from Motor Vehicles behave differently from the other crimes, being concentrated near entertainment districts such as downtown, rather than in traditional "inner-city" areas. Crime patterns generally differ by time of day, with Robberies more likely than others to be reported at night. While proximity to multiunit housing is correlated with crime rates, our regression shows that daytime, but not nighttime, robberies—and no burglaries—are influenced by this measure when controlling for income, density, and housing tenure. Car-related thefts are affected as well. This therefore suggests specific avenues where law-enforcement and community resources can best be applied. Our paper proceeds as follows: Section II outlines the statistical methodology. Section III describes the results. Section IV concludes.

## II. Methodology

Because this study examines acquisitive crime, its proximity to rental housing, and its relationship to other socioeconomic factors, our data come from three sources. First, address data for reported crimes, covering all of 2014, were taken from the City of Milwaukee's website. We focus exclusively on four types of "acquisitive" crime: Burglary, Robbery, Theft of Motor Vehicle, and Theft from Motor Vehicle. These reported crimes were then geocoded by address and both used as point data and applied to their corresponding Census block groups for this analysis.

Because these reports have a time associated with each record, we are able to separate them into “nighttime” and “daytime” crime. Although Boivin and Ouellet (2014) note that the time of crime reports in Canada are often subject to the officers’ discretion, we assume that it is closely connected to the time during which the crime was committed or noticed by the victim. We can then examine differences in reporting time. While there is some leeway as to this precise definition, we opt for a six-hour span that is dark throughout the year. Based on the data (depicted as histograms below), we choose 10pm to 4am for all crimes except Robberies, for which we define “nighttime” as 9pm to 3am.[1] Table 1 provides a summary of these four types of crime. Only Robberies occur disproportionately during the nighttime hours, with more than 25 percent of all incidents occurring during this quarter of the day.[2] Based on these data, we expect to find differing results for different types of crime.

Second, we use parcel data from the City of Milwaukee’s 2014 Master Property Record database. Here, we note Ford (1986), who focuses on purpose-built structures with ten or more units. Of the roughly 160,000 properties in Milwaukee, we choose only what we consider to be “large” multiunit housing. We select all properties coded as 8830 (multi-family residential), as well as 8899 (mixed residential-commercial). We choose a number of units that reflects purpose-built rental housing that is large enough to provide a degree of anonymity; here, this threshold is 24 units or greater. Such buildings generally have multiple floors, as well as a long hallway, which can increase anonymity among residents. But, since a number of smaller buildings can form a complex, we follow Ford’s (1986) definition and include groups of single parcels with more than 10 units that are located less than 30 feet apart from one another.

[1] While the definition of “nighttime” is somewhat arbitrary, the 10pm-4am window is preferred because it is dark year-round in Milwaukee. One option is to create a “moving” window that corresponds to actual sunrise and sunset times, but this might put too much emphasis on the amount of sunlight rather than on other factors. A window beginning at 10pm still shows disproportionate numbers of robberies (more than 30 percent of the total).

[2] The choice of 9pm is made to avoid removing the most distinct “non-arbitrary” pattern in the data. Since moving the windows to match for the other crime would cover daylight hours in summer, they were left unchanged.

In considering these criteria, we also applied certain other thresholds for building size and location, corroborating our present analysis by examining the resulting maps, and by using Google Street View. We confirm that these parcels do indeed represent the type of housing that is the focus of our study. In addition, because parcels themselves do not always perfectly correspond to the buildings that are located on them, we also examined the database's address data, and find that when multiple addresses are located on a single parcel, they typically represent separate units within one building.[3] Our selected parcels, 1,044 in all, are mapped in Figure 1. While somewhat arbitrary (but based on sound reasoning), this choice of parameters results in approximately 1,000 parcels that cover roughly one half of the city (51 percent of the city's land area falls within a quarter mile of one of these parcels).

Finally, we retrieve socioeconomic data for 661 block groups (including population, ethnic characteristics, housing tenure, and income) from the U.S. Census (2014 ACS 5-year estimates). Address data for liquor licenses, taken from the City of Milwaukee's website (for 2016), are used as an additional control variable. These are geocoded using ArcGIS and assigned to their corresponding block groups as well. Since this study incorporates socioeconomic variables, it is conducted at the block-group level, which is the smallest areal unit for which complete data are available.

We then use all our data in our spatial analysis. We first examine crime patterns by type and over time, calculating the median time (to confirm which crimes are more likely to be reported later in the day) and graphing hour-by-hour histograms for each of our four categories. We also calculate distances from each crime to its nearest multiunit parcel. Next, we estimate the degree of spatial autocorrelation among the per-capita crime rate (crimes per 1,000 residents) for each type within each block group, for the entire day as well as during the daytime and nighttime only. This is done using the well-known Moran's *I* statistic:

$$I = \frac{n}{\sum_{i=1}^{n}\sum_{j=1}^{n} w_{ij}} \frac{\sum_{i=1}^{n}\sum_{j=1}^{n} w_{ij}(x_i - \bar{x})(x_j - \bar{x})}{\sum_{i=1}^{n}(x_i - \bar{x})^2} \qquad (2).$$

---

[3] Housing types (such as differentiating between condominiums and rented apartments)is not done in this case; as ownership data are not available.

We use the Queen's case or order one (direct contiguity, including corners) as our spatial weight. While other alternatives are available (particularly inverse distance), we follow LeSage (2014) and choose one that uses a simple weighting matrix.

We then examine associations between crime and our set of socioeconomic variables. We calculate Spearman correlations (a simple, nonparametric measure that is less sensitive to outliers than the parametric Pearson correlation) for both the full-day samples and the night-only samples, and compare differences between the two measures. We also map "hot spots" for all four per capita crime rates within each block group, both daytime and nighttime. This is done using the Getis-Ord (1992, 1995) measure:

$$G_i^* = \frac{\sum_{j}^{n} w_{ij}x_j - \bar{x}\sum_{j}^{n} w_{ij}}{s_x\sqrt{\frac{n\sum_{j}^{n} w_{ij}^2 - \left[\sum_{j}^{n} w_{ij}\right]^2}{n-1}}} \qquad (3).$$

Again, the weight *w* is a Queen contiguity measure of order one. In addition to mapping these clusters' spatial distributions, we provide summary statistics for the crime rates and socioeconomic variables for each set of block groups. This will allow us to compare differences not only between daytime and nighttime hot spots, but also with the city as a whole.

Finally, we estimate a spatial regression model, which incorporates a number of key control variables, as well as a measure of apartment density for each block group. Our model is represented as:

$$crimedens = f\left(liqdens, percrent, percwhite, percvac, popdens, deprivation, QmiParc\right) \quad (4).$$

These variables are based on the previous literature, much of which is described above. Another important determinant of crime is proxied by the density of liquor licenses, which can be both a cause and a product of neighborhood deterioration. Such a connection to crime rates has been studied by Toomey *et al.* (2012) for Minneapolis, Pridemore and Grubescic (2013) for Cincinnati, and Lipton *et al.* (2013) for Boston.

Our explanatory variables include the percentage of renting households; the percentage of white residents; the vacancy rate; population density; and an index of economic deprivation measured as the first principal component of each block group's poverty rate, unemployment rate, percentage of adults above age 25 without a high-school diploma, and percentage of SNAP recipients. The variable *QmiParc* is the number of multi-unit housing parcels within a quarter mile of each block group centroid.[4] According to our theory, we expect that the higher the density of these parcels, the greater the crime rate. We estimate this model for per capita crimes in each category for the whole day, as well as for only daytime and only nighttime crimes.

Our spatial lag model is derived as follows:

$$y = \rho W y + X\beta + \varepsilon \quad (5a),$$

$$\left(I - \rho W\right) y = X\beta + \varepsilon \quad (5b),$$

$$y = \left(I - \rho W\right)^{-1}\left(X\beta + \varepsilon\right) \quad (5c).$$

Much of the (non-spatial) statistical analysis was performed using *Eviews*, and other spatial statistics using ArcGIS, this regression is estimated with *GeoDa*. As before, the weight matrix *W* represents first-order Queen contiguity. Our regression allows us to see whether, even if we control for rental rates and other population and socioeconomic characteristics, these crimes are more likely to occur near large apartment buildings. We can also see whether these effects differ by day and by night. Our results for these and other tests are provided below.

[4] This distance represents a "long" distance in terms of line-of-sight or to walk after parking a car. We also tested other ranges, such as one-half mile.

## III. Results

Table 1 provides summary statistics for all crimes. Theft of Motor Vehicle is the most common type of acquisitive crime, followed by Robbery. Separating out a six-hour "nighttime" period for each type of crime, we see that even though this represents one-fourth of the day, only Robberies register more than 25 percent of reported crimes at night. Theft from Motor Vehicle has the lowest percentages; overall, about 10-15 percent of crimes are reported at night. One important caveat is that reporting times might differ from the time the incidents occurred. Nonetheless, in the absence of more accurate data, these times still allow for certain useful conclusions.[5]

Histograms, showing the number of crimes reported during each hour, are presented in Figure 2. While Robberies see high levels of occurrences at night (prompting us to follow the corresponding "peaks" and use the 9pm to 3am period as "nighttime" for this crime rather than 10pm-4am), Burglaries typically are reported in the middle of the day. The two auto-related acquisitive crimes see large numbers of reports in the mid-morning hours, perhaps reflecting when owners return to their vehicles and realize that a crime has occurred. Figure 2 also provides the median time for each crime, which is the hour and minute that represents the midway point between midnight and midnight before which exactly half of the crimes are reported. Thefts from Motor Vehicles are most likely to be reported early in the day, with Robberies latest. This confirms the particular nature of Robbery reports, which are reported in "real time" at night but are also shown below to exhibit important daytime patterns.

[5] One option is to somehow adjust the windows, so that morning reports might be considered to represent "nighttime" crime, but that is not attempted here.

In our second branch of descriptive analysis, we next examine our selected set of large apartment buildings. Figure 3 presents the density of parcels by distance from the CBD (measured as the address of the U.S. Bank building downtown). Most parcels are located at a distance of around two miles of the center, with small "humps" about seven and 11 miles out. Figure 1 depicts 2-mile, 7-mile, and 11-mile bands; indeed there are a number of parcels on the Far South and far Northwest Sides, both of which were formerly suburban towns that were annexed by Milwaukee in the 1950s. Figure 3 also depicts a histogram of parcel size (number of units); most buildings are relatively small, but there are a few very large ones.

Are acquisitive crimes more likely to occur near these buildings? Using ArcGIS, we estimate that about 51 percent of Milwaukee's land area lies within one quarter mile of one of these parcels. Table 2 shows that disproportionately large percentages of crimes occur within this distance, with small increases at night. Thefts of Motor Vehicles, however, are more likely to be reported nearby during the day. For all crimes and times of day, the median distance from crimes to parcels are less than one quarter mile. Burglaries, on average, occur furthest from these buildings, particularly at night. As we see below, this might be due to the fact that single-family homes present more attractive targets than do the residents of large buildings. Other crimes occur closer to these parcels, and the distance is smaller during the day than at night. This is especially true for Theft from Motor Vehicles, with the smallest median distance for all crimes.

We next examine the spatial distribution of these crimes, as well as their underlying socioeconomic determinants. Figure 4 maps the number of crimes per 1,000 residents for our four types of crime. As might be expected, most are concentrated in the inner city and on the Northwest Side, which is also relatively low-income. Theft from Motor Vehicles, however, frequently occurs downtown. Examining daytime and nighttime crime, we find that while all four types of crime are indeed clustered, nighttime reports tend to be more dispersed. This is shown numerically and graphically. Spatial autocorrelation is shown in Table 3 to be fairly high (above 0.3 in all cases), but the Moran coefficient is lower for nighttime crimes. Getis-Ord "hot spots" (as well as low-crime "cold spots"), for both daytime and nighttime crimes, are presented in Figure 5; nighttime crimes appear to be less concentrated.

Particular patterns and differences between the two time periods also emerge. Crime reports tend to have hot spots on Milwaukee's North side, with the exception of Thefts from Motor Vehicles Clusters also appear to differ by crime and by time of day Robberies are more likely to include clusters in the city's whiter and more college-educated East Side. Car thefts are reported to take place further north from the traditional inner city. Thefts from Motor Vehicles, on the other hand, appear to have similar clusters during both periods of the day.

After examining summary statistics for the entire area in Table 4, and for hot spots in Table 5, we are able to examine differences both in crime (by night and day), and in block-group socioeconomic characteristics. Robberies and Burglaries occur in block groups with median incomes lower than the city average; car-related crimes, on the other hand, occur in block groups that are right around the average income. Daytime robberies and nighttime burglaries are clustered in block groups that are, on average, more deprived than block groups in hot spots corresponding to the other time of day.

Thefts from Motor Vehicles are again concentrated in very different areas from the other types of acquisitive crime. For example, the average white population is much higher in these "hot spots" (both daytime and nighttime) than is the case for the other categories. Likewise, the density of liquor licenses is higher than the citywide average. Similar differences can be shown for the other variables as well. Deprivation is higher than the city mean for all crimes and times of day, except for daytime (but not nighttime) Thefts from Motor Vehicles. The percentage of renters is also higher than the city average in all hot spots except for this one type of crime, which occurs in the main business and entertainment districts.

The key variable in this study is indeed related to these crime variables, although in varying ways. The density of large apartment buildings (*QmiParc*) is higher for Robberies and lower for Burglaries and Thefts of Motor Vehicles. (This density is much higher for Thefts *from* Motor Vehicles, again suggesting that these crimes are concentrated in areas with large numbers of apartments.) Table 6 explores these connections, presenting the Spearman correlations among crimes and all other variables. All crime rates are highly correlated with one another. The density of liquor licenses is most highly correlated with Thefts from Motor Vehicle (supporting our findings regarding their occurrences close to bars downtown and on the East Side); correlations are also high with Robberies. Socioeconomic deprivation is most closely associated with Robbery and car thefts. The associations with deprivation are weakest for Theft from Motor Vehicles, but the correlations are high between these thefts and the number of parcels within a quarter mile. The correlation between apartment buildings and acquisitive crime is lowest for Burglaries, again suggesting that single-family homes might be a more attractive target. Most likely, criminals would target homes or smaller apartment buildings because of the relative ease of accessibility and difficulty of detection compared to larger buildings. In addition, homeowners might present a higher expected payoff to a break-in due to higher wealth.

Table 7 provides correlations for nighttime crimes and socioeconomic variables, as well as the difference between the coefficients between nighttime and all crimes. Nighttime crimes of all types are more closely connected to neighborhoods that are whiter and higher-income and more likely to occur close to downtown. They are also less correlated with liquor licenses, renters, vacancies, and deprivation. There is a slight difference between the correlations between crime and apartment density as well, with Thefts from Motor Vehicles less correlated with this density at night versus during the day.

Do the correlations between rental properties and crime hold, once population and tenure characteristics are controlled for? Table 8 provides the results of our spatial regression model. In general, we find a number of key relationships, but these differ by crime and by time of day. While population density is a significant determinant of all crimes, the density of liquor licenses is positively related to all types of crime during the daytime but only contributes to daytime reports of Burglaries and Thefts from Motor Vehicles. Thefts from Motor Vehicles, again, are different areas, which is shown by the fact that this is the only crime category not to carry a negative coefficient for *Percwhite*. Deprivation carries an insignificant coefficient for almost all cases—perhaps due to the fact that acquisitive crimes occur in high- as well as low-income areas, with only daytime robberies are more likely to occur in deprived areas. This finding, as well as its relationship to Social Disorganization Theory and its implications for community and policing strategy, are worthy of further investigation.

Focusing on the density of large rental properties (*QmiParc*), we find that the number of these parcels within a quarter mile of the block group centroid does indeed contribute to higher crime rates within the block group in three key cases. First, daytime motor vehicle thefts are positively affected by this proximity, as are nighttime thefts from motor vehicles. This suggests that a lack of monitoring and large distances between residents and their parked vehicles might be partially responsible. Second, apartment density has only limited influence on Burglaries, perhaps because other types of housing (particularly single-family homes) might be equally attractive. This variable shows no difference between daytime and nighttime crimes that might have been suggested by earlier studies.

Third, only daytime robberies are positively influenced by this proximity. Potential victims might be unemployed or work hours outside of the traditional "first shift," and their needs might therefore require special attention if crime is to be reduced. In general, programs might be proposed to heighten awareness of or harden targets against potential crime, particularly robberies that occur in poorer neighborhoods during the daytime.

## IV. Conclusion

Because large multiunit housing helps foster a degree of anonymity that can lead to a deterioration in social networks and other forms of disorganization, crime rates might be expected to be higher near such buildings. These rates might vary by time of day, given differences in employment and social activities that leave residences and vehicles unoccupied or bring potential victims near to perpetrators. While the first set of ideas, regarding housing tenure and social disorganization, have been widely examined, less has been done to study differences in crime patterns by time of day.

This study combines both types of analysis, examining multiunit housing and the patterns of four types of "acquisitive" crime in the city of Milwaukee during the year 2014. After first selecting a subset of about 1,000 parcels (of over 160,000 in the city) that meet certain thresholds that classify them as "multiunit residential," we map them and find that about half of the city's area is located within a quarter mile of one of these parcels. We then geocode address data for our four crime types, noting not only their disproportionate proximity to these multiunit parcels, but also their disproportionate reporting during daytime hours. Only for Robberies do more than 25 percent of crimes occur during the six-hour “nighttime” period.

Combining GIS and statistical techniques, we calculate a number of correlations and conduct “hot spot” analysis for both daytime and nighttime crimes. Overall, we find that Thefts from Motor Vehicles are more likely to occur close to one of the selected buildings during the nighttime hours. On the other hand, Thefts of Motor Vehicles are more likely to be reported nearby during the day. In general, Thefts from Motor Vehicles exhibit different patterns from the others, occurring downtown rather than in the traditional inner city.

Our spatial statistical analysis shows nighttime crimes to be more dispersed than daytime crimes, with lower spatial autocorrelation coefficients and “hot spots” that are more spatially separated. Examining the statistical properties of crime “hot spots,” we also see differences in their underlying socioeconomic makeup, with Robberies and Burglaries occurring in block groups with lower median incomes than are auto-related crimes. While these crimes do indeed differ by type and time of day, it is necessary to isolate the role of multiunit housing in their patterns of reporting.

Estimating a spatial regression model that controls for housing tenure, race, economic conditions, social disorganization, and population density, we find that daytime motor vehicle thefts, and nighttime thefts from motor vehicles, are increased by proximity to large multiunit parcels. This might be due to insufficient monitoring of parked vehicles in large complexes. Burglaries, on the other hand, are unaffected by proximity to large multiunit parcels, while daytime robberies are higher closer to these types of property.

These findings can be applied in three key ways. First, because the four categories of acquisitive crime described here show their reports to “peak” during different times of day, further research might be able to explain in more detail why they occur (or are reported) during those times. This is particularly true for Robberies, the only category of crime to be disproportionately reported at night. Second, because proximity to multiunit housing is shown to contribute to daytime robberies and car thefts, efforts can be made to reduce social disorganization in the affected areas. Increased guardianship in these areas (such as neighborhood watch groups) might be effective, as might other programs to keep residents involved in their neighborhood. If these crimes are due to relatively low daytime populations, people might need to be particularly careful in securing their vehicles and taking steps to protect their personal property.

Finally, the block-group-specific results show the areas where each time of crime is most likely to occur, both during the day and at night, as well as the socioeconomic profiles of these neighborhoods. Daytime robberies are worthy of particular attention. In the short run, law enforcement might be well-served to increase patrols in lower-income neighborhoods, while long-run development strategies might help improve their underlying characteristics.

## Table 1: Number of Crimes, 2014, by Time of Day.

| | All | Day | Night | % Night |
|---|---|---|---|---|
| Burglaries | 5884 | 5091 | 793 | 13.48 |
| Robberies | 3581 | 2335 | 1246 | 34.79 |
| Theft of MV | 6620 | 5558 | 1062 | 16.04 |
| Theft From MV | 3853 | 3402 | 451 | 11.71 |

Note: "Night" is defined as 10:00pm to 3:59am, except for Robberies, which is 9:00pm-2:59am.

## Table 2: Distance to Nearest Housing Unit for Crime Types.

| | Number of Crimes Reported Within ¼ Mile | | | Med. Distance (Miles) | | |
|---|---|---|---|---|---|---|
| | All | Day | Night | All | Day | Night |
| Burglaries | 3438 (58.4%) | 2945 (57.8%) | 493 (62.2%) | 0.204 | 0.184 | 0.243 |
| Robberies | 2252 (62.9%) | 1462 (62.6%) | 790 (63.4%) | 0.183 | 0.186 | 0.177 |
| Theft of MV | 4224 (63.8%) | 3555 (64.0%) | 669 (63.0%) | 0.173 | 0.175 | 0.173 |
| Theft From MV | 2626 (68.2%) | 2305 (67.8%) | 321 (71.2%) | 0.145 | 0.151 | 0.112 |

## Table 3: Spatial Autocorrelation (Moran's *I*)

| Crime | All | Day | Night |
|---|---|---|---|
| Burglaries | 0.390 | 0.351 | 0.246 |
| Robberies | 0.428 | 0.410 | 0.268 |
| Theft of MV | 0.441 | 0.417 | 0.292 |
| Theft From MV | 0.308 | 0.313 | 0.193 |

## Table 4: Summary Statistics.

| Variable | Mean | SD | Min | Max |
|---|---|---|---|---|
| Burglaries_All | 9.77 | 8.48 | 0 | 78.20 |
| Burglaries_Night | 1.32 | 1.74 | 0 | 11.30 |
| Burglaries_Day | 8.45 | 7.50 | 0 | 77.37 |
| TheftFromMV_All | 6.08 | 7.83 | 0 | 119.02 |
| TheftFromMV_Night | 0.74 | 2.27 | 0 | 33.37 |
| TheftFromMV_Day | 5.35 | 6.21 | 0 | 85.65 |
| Robberies_All | 6.31 | 7.05 | 0 | 57.03 |
| Robberies_Night | 2.18 | 2.71 | 0 | 19.01 |
| Robberies_Day | 4.13 | 4.96 | 0 | 38.02 |
| TheftOfMV_All | 10.82 | 10.70 | 0 | 106.46 |
| TheftOfMV_Night | 1.74 | 2.20 | 0 | 19.87 |
| TheftOfMV_Day | 9.08 | 9.24 | 0 | 91.25 |
| Liq. Dens | 7.77 | 14.65 | 0 | 154.04 |
| Percrent | 52.77 | 23.80 | 0 | 100 |
| Pop | 1062.70 | 456.44 | 249 | 3215 |
| Percwhite | 51.42 | 35.66 | 0 | 100 |
| Percvac | 10.68 | 9.55 | 0 | 47.27 |
| Popdens | 9965.92 | 6611.66 | 287.42 | 44566.25 |
| MedY | 35596 | 18693 | 8629 | 133929 |
| Deprivation | 0.11 | 1.66 | -2.49 | 4.67 |
| QmiParc | 3.60 | 9.02 | 0 | 59 |

Crimes reported per 1,000 residents.

## Table 5: Summary Statistics For Daytime and Nighttime Crime Hot Spots.

| | Robberies | | | | Burglaries | | | |
|---|---|---|---|---|---|---|---|---|
| | Day | | Night | | Day | | Night | |
| | Mean | SD | Mean | SD | Mean | SD | Mean | SD |
| Deprivation | 1.58 | 1.26 | 1.22 | 1.43 | 1.07 | 1.34 | 1.30 | 1.36 |
| QmiParc | 4.38 | 10.15 | 5.77 | 12.10 | 1.54 | 3.33 | 1.32 | 2.03 |
| Liq. Dens | 7.70 | 8.62 | 8.06 | 10.52 | 4.54 | 6.43 | 5.31 | 7.11 |
| Percwhite | 12.86 | 14.42 | 20.07 | 25.45 | 13.86 | 15.41 | 12.13 | 13.34 |
| Percrent | 65.38 | 18.61 | 65.28 | 18.45 | 59.84 | 18.05 | 61.19 | 17.09 |
| MedY | 25693.28 | 8859.25 | 27283.66 | 11677.39 | 29099.00 | 10020.03 | 27712.24 | 8872.32 |
| | Theft From MV | | | | Theft of MV | | | |
| | Day | | Night | | Day | | Night | |
| | Mean | SD | Mean | SD | Mean | SD | Mean | SD |
| Deprivation | 0.00 | 1.75 | 0.32 | 1.82 | 0.46 | 1.63 | 0.50 | 1.47 |
| QmiParc | 14.71 | 17.36 | 17.37 | 18.44 | 0.41 | 1.06 | 1.45 | 1.82 |
| Liq. Dens | 28.55 | 32.21 | 24.56 | 29.95 | 4.38 | 7.35 | 3.60 | 6.38 |
| Percwhite | 59.62 | 30.22 | 55.66 | 29.27 | 24.26 | 28.93 | 21.83 | 22.43 |
| Percrent | 74.99 | 15.60 | 74.75 | 19.94 | 49.64 | 22.18 | 51.87 | 20.92 |
| MedY | 37644.40 | 18833.14 | 34081.09 | 21625.51 | 35384.01 | 15116.31 | 34347.57 | 12715.59 |

## Table 6: Spearman Correlations Among (All) Crimes and Other Variables.

| | BURG | TFMV | ROB | TMV | LIQDENS | PERCRENT | PERCWH | PERCVAC | POPDENS | LNMEDY | DEPR | QMIPARC |
|---|---|---|---|---|---|---|---|---|---|---|---|---|
| TFMV | 0.48 | 1 | | | | | | | | | | |
| ROB | 0.69 | 0.52 | 1 | | | | | | | | | |
| TMV | 0.69 | 0.49 | 0.72 | 1 | | | | | | | | |
| LIQDENS | 0.16 | 0.37 | 0.35 | 0.16 | 1 | | | | | | | |
| PERCRENT | 0.25 | 0.28 | 0.47 | 0.35 | 0.37 | 1 | | | | | | |
| PERCWHITE | -0.63 | -0.24 | -0.66 | -0.65 | -0.05 | -0.41 | 1 | | | | | |
| PERCVAC | 0.36 | 0.24 | 0.45 | 0.33 | 0.22 | 0.37 | -0.44 | 1 | | | | |
| POPDENS | 0.09 | 0.03 | 0.19 | 0.08 | 0.38 | 0.38 | -0.16 | 0.17 | 1 | | | |
| LNMEDY | -0.39 | -0.24 | -0.57 | -0.46 | -0.28 | -0.69 | 0.60 | -0.44 | -0.30 | 1 | | |
| DEPRIVATION | 0.45 | 0.20 | 0.62 | 0.51 | 0.27 | 0.61 | -0.69 | 0.45 | 0.34 | -0.85 | 1 | |
| QMIPARC | 0.01 | 0.21 | 0.16 | 0.13 | 0.22 | 0.45 | -0.02 | 0.13 | 0.20 | -0.28 | 0.16 | 1 |
| LNDISTCBD | -0.12 | -0.35 | -0.37 | -0.06 | -0.57 | -0.49 | 0.07 | -0.32 | -0.49 | 0.37 | -0.35 | -0.32 |

# Table 7: Correlations For Nighttime Crimes And Differences With Full-Sample Correlations.

| | Spearman | | | | Difference by which Corr Nite> Corr All | | | |
|---|---|---|---|---|---|---|---|---|
| Variable | BURG | TFMV | ROB | TMV | BURGN | TFMVN | ROBN | TMVN |
| TFMV | 0.11 | 1.00 | | | -0.37 | | | |
| ROB | 0.41 | 0.21 | 1.00 | | -0.28 | -0.30 | | |
| TMV | 0.36 | 0.16 | 0.46 | 1.00 | -0.33 | -0.33 | -0.26 | |
| LIQDENS | 0.08 | 0.25 | 0.31 | 0.10 | -0.08 | -0.12 | -0.04 | -0.06 |
| PERCRENT | 0.20 | 0.20 | 0.41 | 0.25 | -0.05 | -0.07 | -0.06 | -0.10 |
| PERCWHITE | -0.48 | -0.12 | -0.53 | -0.52 | 0.14 | 0.12 | 0.13 | 0.13 |
| PERCVAC | 0.28 | 0.09 | 0.34 | 0.20 | -0.08 | -0.15 | -0.11 | -0.13 |
| POPDENS | 0.05 | 0.00 | 0.20 | 0.09 | -0.04 | -0.03 | 0.01 | 0.01 |
| LNMEDY | -0.32 | -0.17 | -0.48 | -0.35 | 0.06 | 0.07 | 0.09 | 0.11 |
| DEPR | 0.36 | 0.14 | 0.49 | 0.40 | -0.09 | -0.07 | -0.12 | -0.11 |
| QMIPARC | -0.01 | 0.12 | 0.16 | 0.07 | -0.02 | -0.08 | -0.01 | -0.06 |
| LNDISTCBD | 0.00 | -0.22 | -0.32 | -0.03 | 0.12 | 0.12 | 0.05 | 0.03 |

# Table 8: Regression Results (p-values in parentheses).

| | MV Theft | | | Burglaries | | |
|---|---|---|---|---|---|---|
| | All | Night | Day | All | Night | Day |
| ρ | **0.41 (0.000)** | **0.30 (0.000)** | **0.39 (0.000)** | **0.29 (0.000)** | 0.09 (0.119) | **0.28 (0.000)** |
| CONSTANT | **12.66 (0.000)** | **2.57 (0.000)** | **10.77 (0.000)** | **12.29 (0.000)** | **2.06 (0.000)** | **10.85 (0.000)** |
| LIQDENS | **0.08 (0.002)** | 0.01 (0.054) | **0.07 (0.003)** | **0.06 (0.003)** | 0.00 (0.735) | **0.06 (0.002)** |
| PERCRENT | 0.00 (0.904) | 0.00 (0.658) | 0.00 (0.801) | -0.02 (0.185) | 0.00 (0.780) | -0.02 (0.129) |
| PERCWHITE | **-0.09 (0.000)** | **-0.02 (0.000)** | **-0.07 (0.000)** | **-0.09 (0.000)** | **-0.02 (0.000)** | **-0.07 (0.000)** |
| PERCVAC | 0.02 (0.616) | 0.00 (0.958) | 0.02 (0.572) | **0.13 (0.000)** | **0.03 (0.001)** | **0.11 (0.000)** |
| DEPRIVATION | 0.51 (0.120) | 0.05 (0.483) | 0.44 (0.131) | 0.11 (0.687) | 0.02 (0.803) | 0.07 (0.760) |
| POPDENS | **0.00 (0.000)** | **0.00 (0.003)** | **0.00 (0.000)** | **0.00 (0.000)** | **0.00 (0.004)** | **0.00 (0.000)** |
| QMIPARC | 0.09 (0.056) | 0.01 (0.463) | **0.08 (0.047)** | 0.03 (0.434) | 0.00 (0.811) | 0.03 (0.445) |
| $R^2$ | 0.41 | 0.24 | 0.39 | 0.39 | 0.25 | 0.35 |
| AIC | 4486.01 | 2637.43 | 4323.22 | 4201.34 | 2320.52 | 4089.02 |
| | Robberies | | | Theft From MV | | |
| | All | Night | Day | All | Night | Day |
| ρ | **0.30 (0.000)** | **0.19 (0.001)** | **0.23 (0.000)** | **0.43 (0.000)** | **0.32 (0.000)** | **0.43 (0.000)** |
| CONSTANT | **7.23 (0.000)** | **2.39 (0.000)** | **5.56 (0.000)** | **4.13 (0.000)** | 0.64 (0.064) | **3.55 (0.000)** |
| LIQDENS | **0.10 (0.000)** | **0.04 (0.000)** | **0.06 (0.000)** | **0.20 (0.000)** | **0.04 (0.000)** | **0.16 (0.000)** |
| PERCRENT | 0.00 (0.902) | 0.01 (0.331) | 0.00 (0.688) | 0.02 (0.191) | 0.01 (0.141) | 0.01 (0.261) |
| PERCWHITE | **-0.05 (0.000)** | **-0.02 (0.000)** | **-0.04 (0.000)** | -0.01 (0.324) | 0.00 (0.998) | -0.01 (0.212) |
| PERCVAC | **0.11 (0.000)** | **0.04 (0.001)** | **0.08 (0.000)** | 0.04 (0.188) | 0.00 (0.664) | 0.05 (0.061) |
| DEPRIVATION | **0.69 (0.001)** | 0.09 (0.308) | **0.62 (0.000)** | 0.12 (0.645) | 0.08 (0.332) | 0.04 (0.853) |
| POPDENS | **0.00 (0.000)** | **0.00 (0.000)** | **0.00 (0.000)** | **0.00 (0.000)** | **0.00 (0.000)** | **0.00 (0.000)** |
| QMIPARC | **0.07 (0.016)** | 0.02 (0.077) | **0.05 (0.015)** | **0.07 (0.046)** | **0.02 (0.030)** | 0.05 (0.067) |
| $R^2$ | 0.47 | 0.30 | 0.45 | 0.37 | 0.21 | 0.36 |
| AIC | 3883.66 | 2847.23 | 3456.02 | 4140.46 | 2704.23 | 3851.38 |

Bold = significant at 5 percent.

**Figure 1: Selected Multiunit Housing (White Polygons) and Median Income in Milwaukee.**

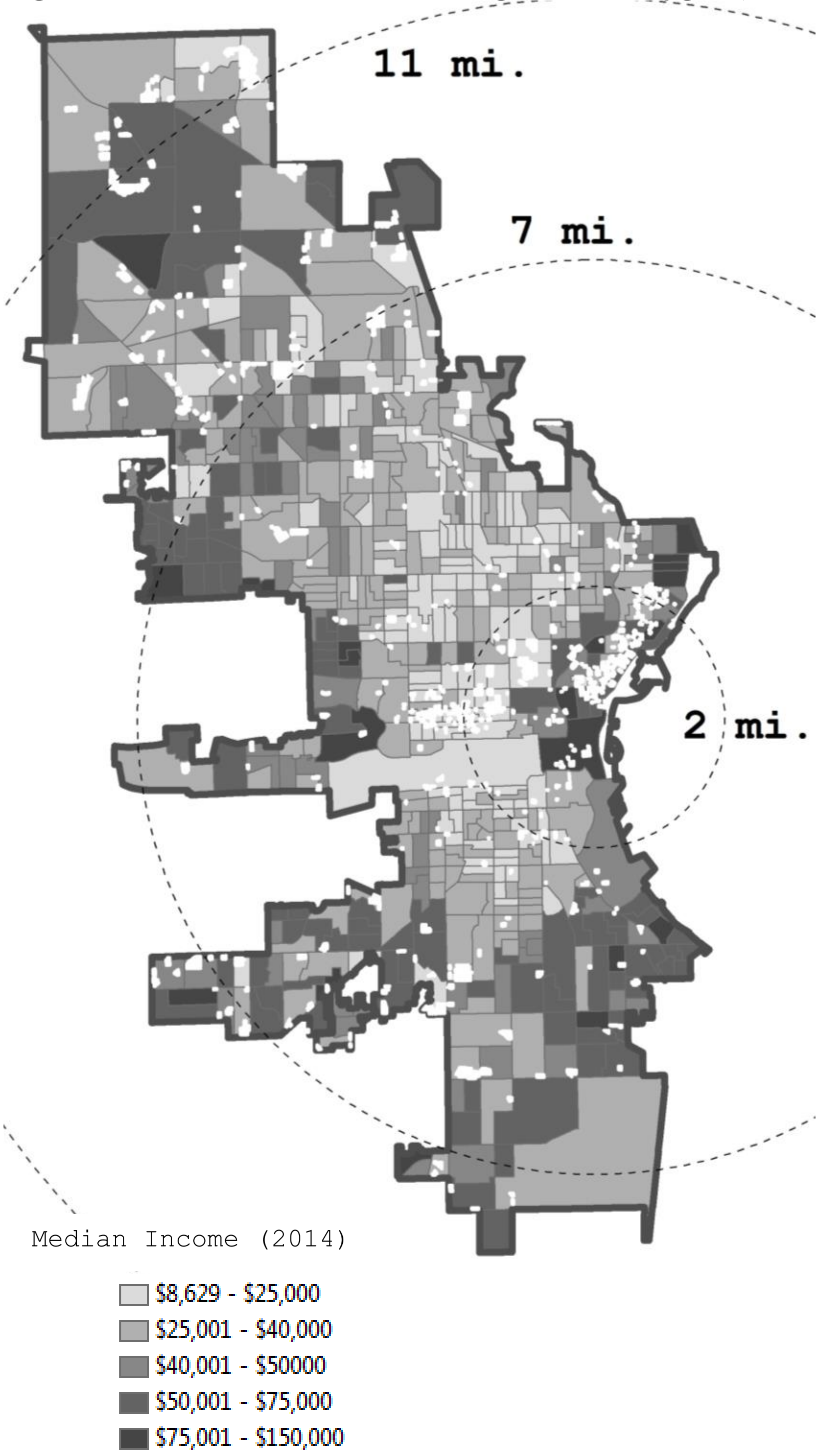

## Figure 2: Number of Crimes By Reporting Hour.

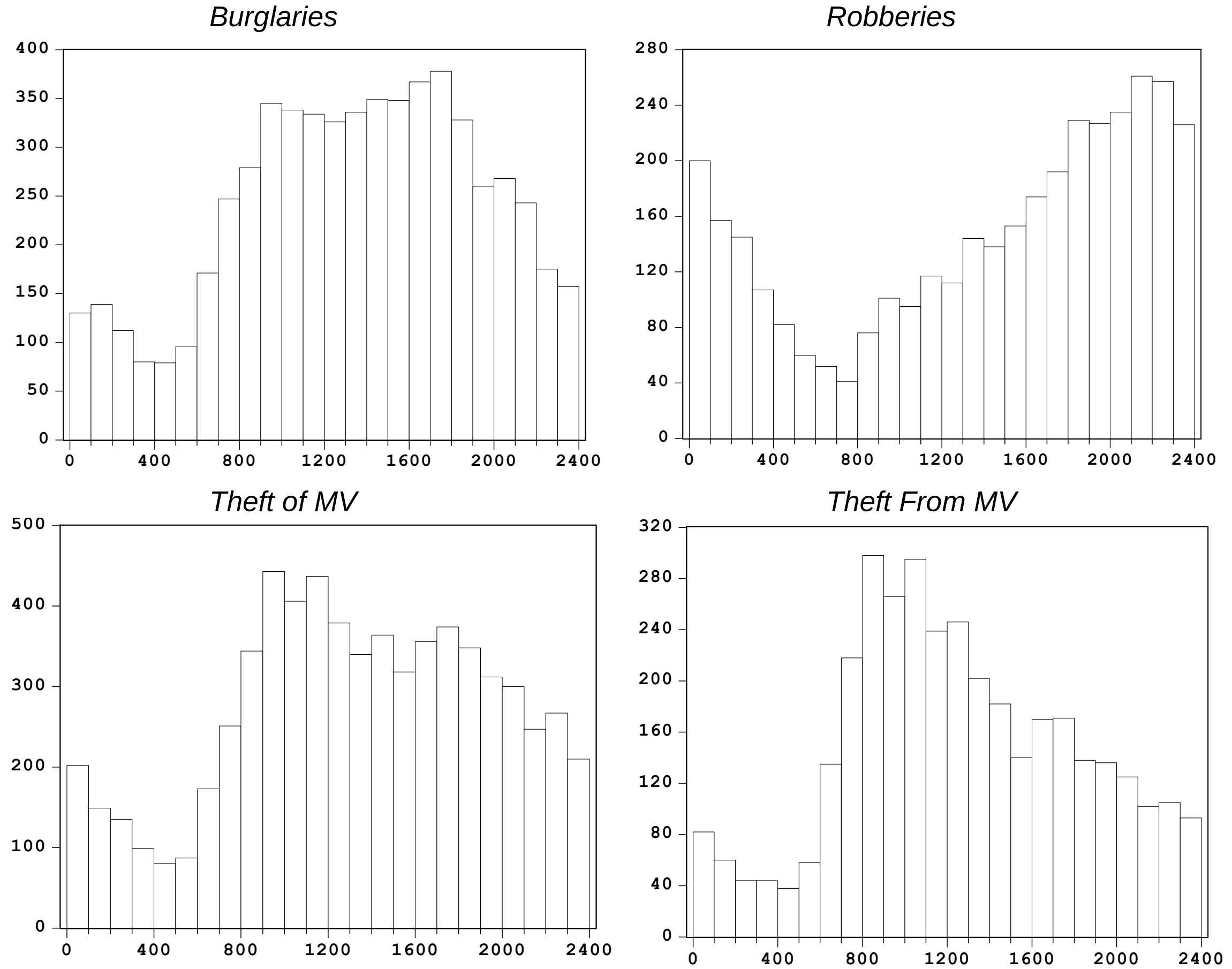

Median time: Burglaries 1:46 PM Robberies 4:02 PM Theft of MV 1:20 PM Theft From MV 12:04 PM

## Figure 3: Distribution of Selected Multiunit Housing in Milwaukee.

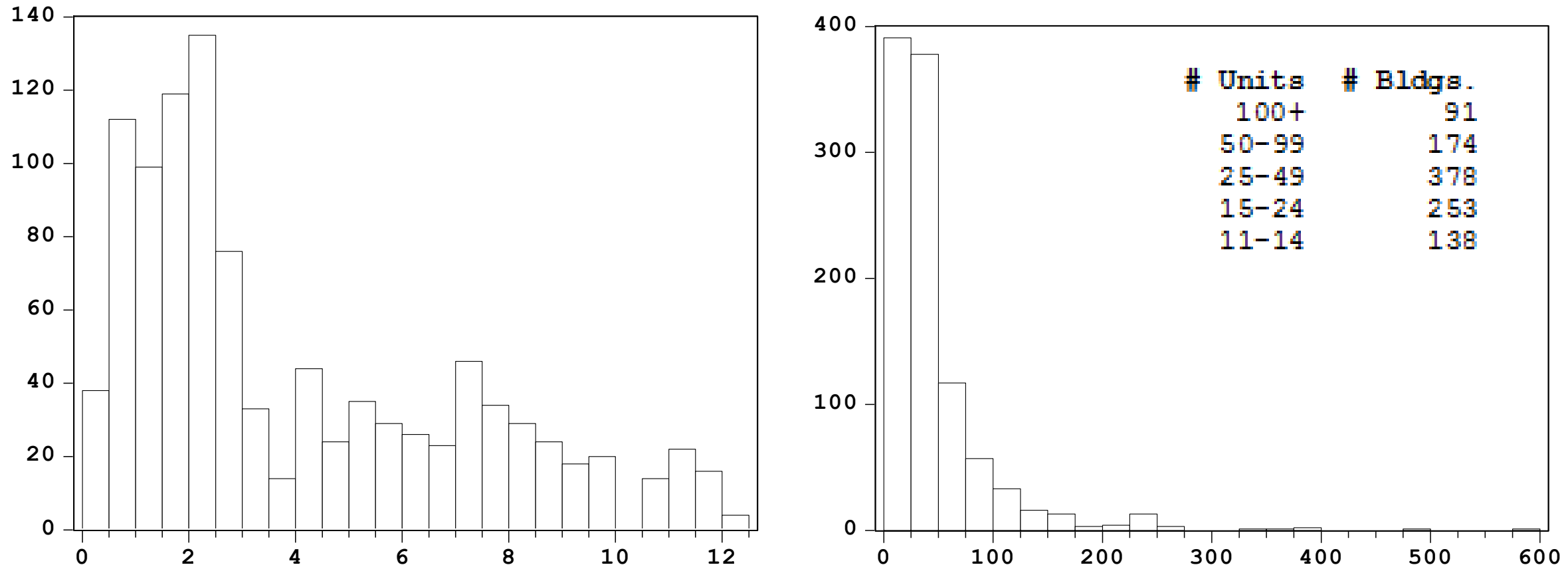

```
# of Buildings (vert.) by Distance from CBD
N = 1034       Median Distance = 2.54 mi
```

```
# of Buildings (vert.) by Units in Building
N = 1034       Median # of Units = 30
```

**Figure 4: Crimes Per 1,000 Inhabitants.**

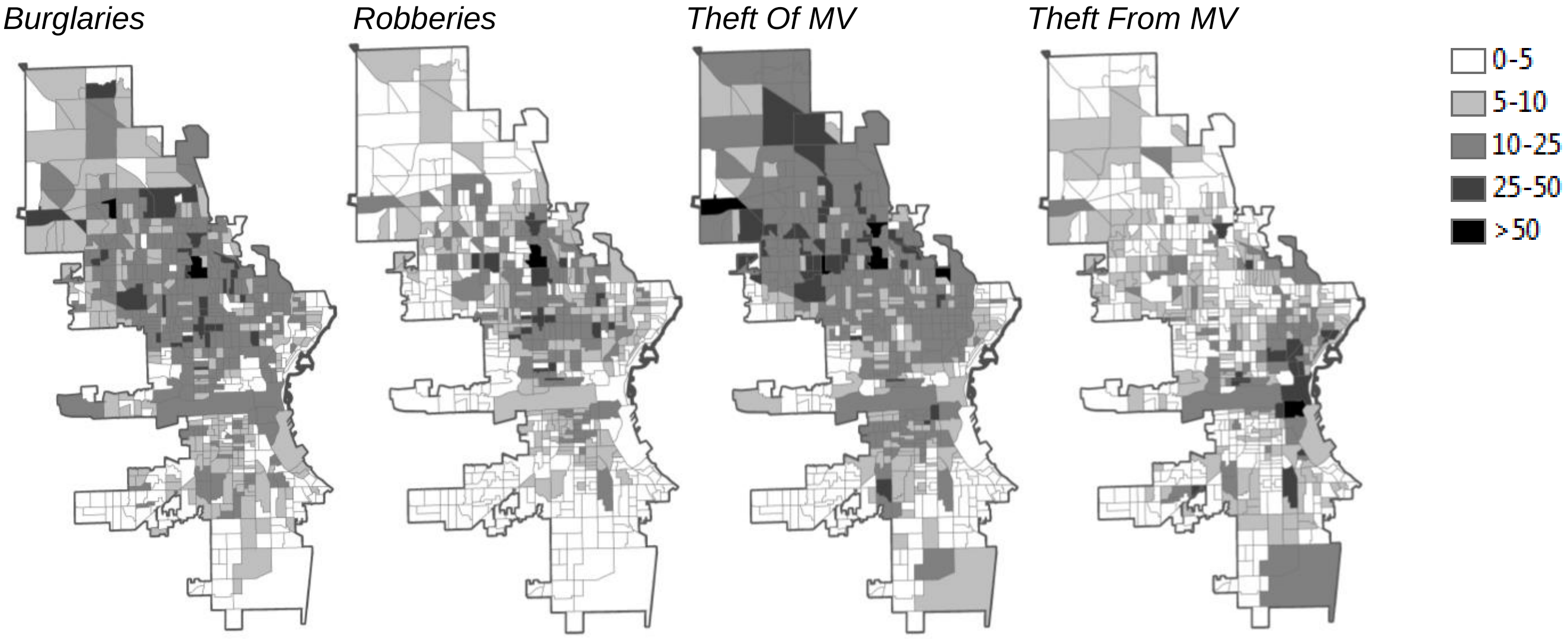

**Figure 5: Getis-Ord "Hot Spots" By Time of Day.**

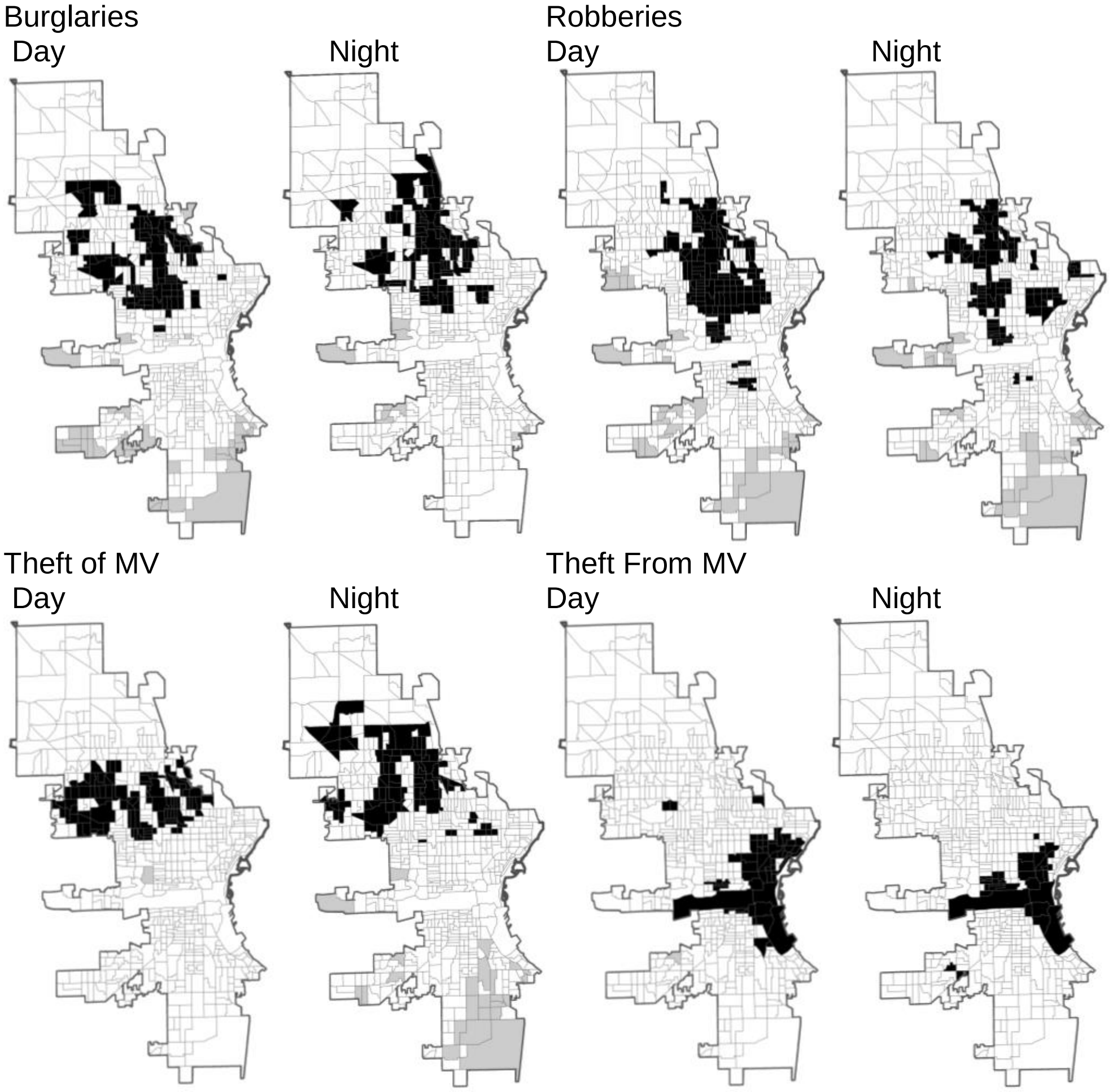